\documentclass[prd,aps,eqsecnum,floatfix,nofootinbib,preprint,tightenlines]{revtex4}

\usepackage{latexsym}
\usepackage{graphicx}
\usepackage{colordvi}
\usepackage{xcolor}
\usepackage{amssymb}

\def\bibi{\bibitem}

\def\d{\delta}
\def\g{\gamma}

\def\k{\kappa}
\def\l{\lambda}
\def\m{\mu}
\def\n{\nu}

\def\p{\pi}                     % Also, \varpi
\def\r{\rho}                    %       \varrho
\def\s{\sigma}                  %       \varsigma

\def\P{\Pi}

\def\cbo{{\,\raise-.15ex\Sc [\,}}                       % curly "
\def\gtap{\raisebox{-.4ex}{\rlap{$\sim$}} \raisebox{.4ex}{$>$}}   % > or ~
\def\ddt#1{{\buildrel {\hbox{\LARGE .\kern-2pt.}} \over {#1}}}% double dot-over
\def\ie{\mbox{\it i.e.}}
\def\eg{\mbox{\it e.g.}}
\def\etc{\mbox{\it etc.}}

\def\half{{1\over 2}}

\def\floatcaption#1#2{ \caption{ #2 \ [#1] \label{#1}} }
\def\floatcaption#1#2{ \caption{#2 \label{#1}} }

\def\ttl#1{{\it #1}}
\def\ttl#1{}

\long\def\symbolfootnote[#1]#2{\begingroup%
\def\thefootnote{\fnsymbol{footnote}}\footnote[#1]{#2}\endgroup}

\long \def \blockcomment #1\endcomment{}

\def\eg{{\it e.g.}}
\def\etc{{\it etc.}}
\def\seef{{\it cf.}}

\def\hq{{\hat q}}
\def\bP{{\overline \P}}
\def\qb{{\overline q}}
\def\htheta{{\hat\theta}}
\newcommand{\ubar}{{\overline{u}}}
\newcommand{\dbar}{{\overline{d}}}

\begin{document}

\begin{center}
\begin{boldmath}
{\large\bf Finite-volume effects in the muon anomalous magnetic moment
on the lattice}\\[0.4cm]
\end{boldmath}
\vspace{3ex}
{Christopher~Aubin,$^a$ Thomas~Blum,$^b$  Peter~Chau,$^c$
Maarten~Golterman,$^c$  
Santiago~Peris,$^d$ Cheng~Tu$^b$
\\[0.1cm]
{\it
\null$^a$Department of Physics \& Engineering Physics\\ Fordham University, Bronx,
New York, NY 10458, USA\\
\null$^b$Physics Department\\
University of Connecticut, Storrs, CT 06269, USA\\
\null$^c$Department of Physics and Astronomy, San Francisco State University\\
San Francisco, CA 94132, USA\\
\null$^d$Department of Physics and IFAE-BIST, Universitat Aut\`onoma de Barcelona\\
E-08193 Bellaterra, Barcelona, Spain}}
\\[6mm]
{ABSTRACT}
\\[2mm]
\end{center}
\begin{quotation} 
We investigate finite-volume effects in the hadronic vacuum polarization, with an eye toward the corresponding systematic error in the muon anomalous magnetic moment.
We consider both recent lattice data as well
as lowest-order, finite-volume chiral perturbation theory, in order to get a quantitative
understanding.
Even though leading-order chiral perturbation theory does not provide a good description of the hadronic vacuum polarization, it turns out that it gives a 
good representation of finite-volume effects. We find that finite-volume effects cannot be ignored when the aim is a few percent level accuracy for the leading-order
hadronic contribution to the muon anomalous magnetic moment, even when 
using ensembles with $m_\pi L\, \gtap\, 4$ and $m_\pi \sim 200$ MeV.
\end{quotation}

\vfill
\eject
\setcounter{footnote}{0}

%%####%%
%\newpage
\section{\label{introduction} Introduction}
%%####%%
A convenient representation for the lowest-order hadronic contribution to the anomalous magnetic moment $a_\m=(g-2)/2$ in terms of an
integral over euclidean momentum is given by \cite{TB2003,ER}
\begin{eqnarray}
\label{amu}
a_\m^{\rm LO,HVP}&=&\lim_{q_{max}^2\to\infty} a_\m^{\rm LO,HVP}[q_{max}^2]\ ,\\
a_\m^{\rm LO,HVP}[q_{max}^2]&=&4\alpha^2\int_0^{q_{max}^2}\!\!\!\!\! dq^2\,f(q^2)\,
{\hat{\Pi}}(q^2)\ ,\nonumber 
\end{eqnarray}
where $m_\m$ is the muon mass,
\begin{eqnarray}
\label{integral}
f(q^2)&\!\!\!\!=&\!\!m_\mu^2 q^2 Z^3(q^2)\,\frac{1-q^2 Z(q^2)}{1+m_\m^2 q^2
  Z^2(q^2)}\ ,\nonumber\\
Z(q^2)&\!\!\!\!=&\!\!\!\left(\sqrt{(q^2)^2+4m_\m^2 q^2}-q^2\right)/
(2m_\m^2 q^2)\ ,
\end{eqnarray}
and $\hat{\Pi}(q^2)\equiv\Pi (q^2)-\Pi (0)$ is the subtracted 
hadronic vacuum polarization. The vacuum polarization $\P(q^2)$ is defined from the hadronic electromagnetic current two-point function, $\P_{\m\n}(q)$, via
\begin{equation}
\label{polndefn}
\P_{\m\n}(q)\equiv
\int d^4x\ e^{iqx} \langle J_\m(x) J_\n(0)\rangle
=\left(q^2\d_{\m\n}-q_\mu q_\nu\right)\P(q^2)\ ,
\end{equation}
with $J_\mu(x)$ the hadronic electromagnetic current.   The form on the
right-hand side of Eq.~(\ref{polndefn}) follows from current conservation and rotational symmetry.

While, in principle, a lattice computation of the hadronic vacuum polarization\footnote{Or, at least, its connected part.} is straightforward, it turns out that a
very high accuracy is needed in the region around $q^2\sim m_\m^2/4$.   The
reason is that the integrand of Eq.~(\ref{amu}) is strongly peaked in that region,
as illustrated in Fig.~\ref{integrandnew}.   In effect, the weight function $f(q^2)$
acts as a ``magnifying glass'' of the low-momentum region, where it is
hard to obtain lattice data points with small errors.   We note that the data
points shown in Fig.~\ref{integrandnew} have been obtained with 
all-mode averaging (AMA) \cite{AMA}, and thus have errors much reduced
in comparison with the state of the art of only a few years ago.\footnote{See,
for example, Fig.~1 of Ref.~\cite{taumodel}.}

Figure~\ref{integrandnew} also suggests that finite-volume effects may cause
a significant systematic error, because it is the finite-volume quantization of
momenta that makes the number of data points in the low-$q^2$ region so
sparse.   It is our aim in this article to investigate this quantitatively.   A more
phenomenological study of finite-volume effects appeared in Ref.~\cite{Mainz},
and a preliminary account of the present work appeared in Ref.~\cite{lat15}.%
\footnote{The results reported in Ref.~\cite{lat15} were based on an
incorrect version of Eq.~(\ref{chptresult}), and did not take into account taste splittings.
}

We will restrict ourselves to a rectangular volume $L^3\times T$ with 
periodic boundary conditions in all directions, and we will be interested in the
case that $T>L$, as is the case for all lattice computations of the hadronic
vacuum polarization.   While twisted boundary conditions have been 
considered \cite{DJJW2012,ABGP13}, a generic twist vector reduces the symmetry group of the problem.
Thus, the continuum representation of the rotation group according to which
$\P_{\m\n}(q)$ transforms would reduce to even more representations of the
even smaller discrete subgroup, but it would not reduce the size of
finite-volume effects given certain values of $m_\p L$ and $m_\p$.

%%%%%%%%%%%%%%%%%%%
\begin{figure}[t]
%\vspace*{4ex}
\begin{center}
\includegraphics*[width=12cm]{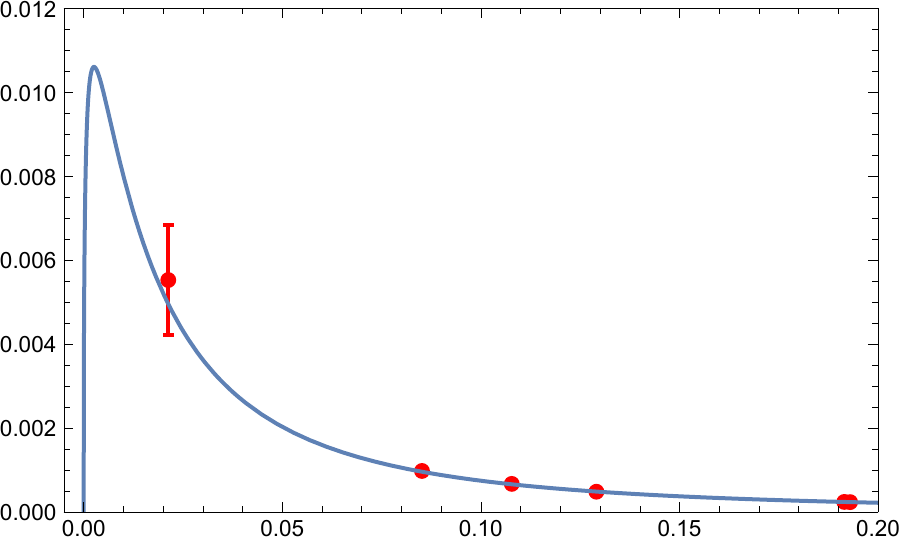}
\end{center}
\vspace*{-6ex}
\begin{quotation}
\floatcaption{integrandnew}%
{{\it Integrand of Eq.~(\ref{amu}), in arbitrary units ($q^2$ in GeV$^2$).   The red
points represent lattice data from the MILC asqtad ensemble considered in this
article, while the blue curve is the product of the weight $f(q^2)$ and a typical
smooth fit to the subtracted vacuum polarization ${\hat{\Pi}}(q^2)$.
}}
\end{quotation}
\vspace*{-4ex}
\end{figure}
%%%%%%%%%%%%%%%%%%%
This article is organized as follows.   In Sec.~\ref{theory} we discuss general
theoretical considerations based on the Ward--Takahashi identity (WTI) and
the irreducible representations of the cubic group, followed by a calculation
of the vacuum polarization in finite volume in lowest-order (staggered)
chiral perturbation theory (ChPT).   In Sec.~\ref{lattice}, we then compare
lattice data for the vacuum polarization with ChPT, and quantify the 
size of the systematic error due to finite-volume effects on 
$a_\m^{\rm LO,HVP}$.   We conclude in Sec.~\ref{conclusion}, and an
appendix contains details of the calculation of the finite-volume vacuum
polarization in ChPT (generalizing it to the case with twisted boundary
conditions).

%%####%%
%\newpage
\section{\label{theory} Theoretical considerations}
%%####%%
In an infinite volume and in the continuum limit, symmetry and current conservation implies
that the vacuum polarization takes the form~(\ref{polndefn}).   Current conservation
carries over to the lattice, but now a more general decomposition of the vacuum polarization is possible, because the
symmetry is reduced.   
The WTI restricts $\P_{\m\n}(q)$ to obey ($a$ is the lattice spacing)
\begin{eqnarray}
\label{WTI}
\sum_\m \hq_\m\P_{\m\n}(q)&=&0\ ,\\
\label{qhat}
\hq_\m&\equiv &\frac{2}{a}\sin{(aq_\m/2)}\ .
\end{eqnarray}
Requiring $\P_{\m\n}(q)$ to be symmetric in the indices $\m$ and $\n$,\footnote{We will always use only the Noether current in Eq.~(\ref{polndefn}).} and assuming
an infinite, isotropic hypercubic lattice, the WTI implies the most general form\footnote{See also Ref.~\cite{JLQCD}.}
\begin{eqnarray}
\label{vacpolanonzero}
\P_{\m\n}(q)&=&\left(\d_{\m\n}\hq^2-\hq_\m \hq_\n\right)\P(q)\\
&&+\left(\d_{\m\n}\left(\sum_\r\hq_\r^4+\hq_\m^2\hq^2\right)
-\hq_\m^3\hq_\n-\hq_\m\hq_\n^3\right)\P'(q)+\dots\ ,\nonumber
\end{eqnarray}
where $\hq^2=\sum_\m\hq_\m^2$.   While $\P(q)$ is dimensionless,
$\P'(q)$ has mass dimension -2.   That means that it has to vanish at least 
quadratically with the lattice spacing $a$; for $a\to 0$, the expression in Eq.~(\ref{vacpolanonzero}) has to reduce to
Eq.~(\ref{polndefn}).   Here, we will be only interested in the vacuum polarization for 
very small momenta, and we will thus assume that we can ignore the scaling violations on the second line of Eq.~(\ref{vacpolanonzero}).

When we restrict ourselves also to a finite volume in the form of a hypercubic
box of dimensions $L^3\times T$ with periodic boundary conditions, Eq.~(\ref{vacpolanonzero})  is not the most general form
of $\P_{\m\n}(q)$ if the hypercubic symmetry is further broken by 
choosing $L\ne T$, as we will
discuss next.

%%####%%
%\newpage
\subsection{\label{group} Group theory}
%%####%%
When we go to a finite periodic volume $L^3\times T$ with $L\ne T$,\footnote{We will always consider the case that $T>L$.} two things happen.
First, momenta are quantized,
\begin{equation}
\label{momquant}
q_i = \frac{2\p n_i}{L}\ ,\quad i=1,2,3\ ,\qquad
q_4= \frac{2\p n_4}{T}\ ,
\end{equation}
where the $n_\m$ are integers.  The WTI~(\ref{WTI}) does not restrict the vacuum
polarization at zero momentum, and in general, in a finite volume, $\P_{\mu\n}(0)\ne 0$.\footnote{This, and some of the other observations that follow below, have also
been noted in Ref.~\cite{BM}.}   Rather, rotational symmetry implies that it takes the
form
\begin{equation}
\label{formPi0}
\P_{\mu\n}(0)=\d_{\m\n}\left(\P_s(0) + \d_{\m 4} \left(\P_4(0)-\P_s(0)\right)\right)\ ,
\end{equation}
with $\P_s(0)$ and $\P_4(0)$ constants.\footnote{For an estimate using ChPT,
see the appendix.}  For $T\gg L$ one expects that $\P_4(0)\ll \P_s(0)$.
It thus makes sense to consider a subtracted vacuum polarization
\begin{eqnarray}
\label{vacpolproj}
\bP_{\m\n}(q)&=&\sum_{\k\l}P^T_{\m\k}\left(\P_{\k\l}(q)-\P_{\k\l}(0)\right)P^T_{\l\n}\\
&=&\P_{\m\n}(q)-P^T_{\m\n}\P_s(0) -
P^T_{\m 4}P^T_{4\n}\left( \P_4(0)-\P_s(0)\right)
\ ,\nonumber\\
P^T_{\m\n}(q)&=&\d_{\m\n}-\frac{q_\m q_\n}{q^2}\ ,\nonumber
\end{eqnarray}
where $P^T$ is the transverse projector.  We projected the subtracted vacuum 
polarization so that it still satisfies the WTI after the subtraction of its value at 
zero momentum.   Of course, without the subtraction, this projection has no 
effect, since $\sum_\m q_\m\P_{\m\n}(q)=\sum_\n \P_{\m\n}(q)q_\n=0$.

Second, the rotation group is reduced to the group of cubic rotations, defined
by the irreducible representation (irrep) of $90$-degree rotations on the spatial
components of momentum.   While the infinite-volume form~(\ref{polndefn})
contains only one scalar function, this is no longer the case in our finite volume.
The tensor $\bP_{\m\n}$ contains five different irreducible
sub-structures:
\begin{equation}
\label{irreps}
\phantom{q_4\sum_i q_i^2\left(\bP_{T_1}-\bP_{A_1^{44}}\right)=0}
\end{equation}
\vspace{-9.3ex}
\begin{center}
\begin{tabular}{ll}
$A_1$: & $ \sum_i \bP_{ii}=(3q^2-{\vec q}\,^2)\bP_{A_1}\ ,$ \\
$A_1^{44}$: & $\bP_{44}=({\vec q}\,^2)\bP_{A_1^{44}}\ ,$ \\
$T_1$: & $\bP_{4i}=-(q_4 q_i)\bP_{T_1}\ ,$\\
$T_2$: & $\bP_{ij}=-(q_i q_j)\bP_{T_2}\ ,\ i\ne j\ ,$\\
$E$: & $\bP_{ii}-\sum_i \bP_{ii}/3=(-q_i^2+{\vec q}\,^2/3)\bP_E\ ,$ \\
\end{tabular}
\end{center}
where ${\vec q}\,^2=\sum_i q_i^2$.   Equation~(\ref{irreps})
defines five different scalar functions $\bP_r$, $r\in\{A_1,A_1^{44},T_1,T_2,E\}$,
which are unrelated by symmetry, since the sub-structures shown here do not
transform into each other under cubic rotations.
For the spatial diagonal elements, Eq.~(\ref{irreps}) implies that
\begin{equation}
\label{spatialdiag}
\bP_{ii}=(-q_i^2+{\vec q}\,^2/3)\bP_E+(q^2-{\vec q}\,^2/3)\bP_{A_1}\ .
\end{equation}
The irrep $A_1$ occurs twice in Eq.~(\ref{irreps}), and we distinguish the two with the notation
$A_1$ and $A_1^{44}$.   
Unbarred scalar functions, $\P_r$, $r\in\{A_1,A_1^{44},T_1,T_2,E\}$, are defined
analogously
by replacing components of $\bP_{\m\n}$ by $\P_{\m\n}$ on the left-hand side
of Eq.~(\ref{irreps}).
The WTI implies some relations between these
functions.   In particular, $\sum_\m q_\m\bP_{\m 4}=0$ implies that
\begin{equation}
\label{WTI1}
q_4{\vec q}\,^2\left(\bP_{T_1}-\bP_{A_1^{44}}\right)=0\ ,
\end{equation}
while $\sum_\m q_\m\bP_{\m i}=0$ implies (choosing $i$ such that $q_i\ne 0$)
\begin{equation}
\label{WTI2}
{\vec q}\,^2\,(-\bP_{T_2}+\frac{1}{3}\,\bP_E+\frac{2}{3}\,\bP_{A_1})
+q_i^2(\bP_{T_2}-\bP_E)+q_4^2\,(\bP_{A_1}-\bP_{T_1})=0\ .
\end{equation}
The unbarred $\P_r$ satisfy the same relations.

We note that these scalar functions can still be functions of all cubic invariants
that can be made out of $q_\m$.    Invariants with dimension larger than 2,
like $\sum_i q_i^4$ have coefficients that are positive power of the lattice
spacing, so we will assume that these are negligibly small at the values of $q_\m$
we are interested in.   That still leaves us with the invariants ${\vec q}\,^2$
and $q_4^2$.\footnote{Odd powers of $q_4$ are excluded because of axis
inversion symmetry.}
Empirically, we find, however, that the functions $\bP_r$ are a function of
$q^2$ (or $\hq^2$, see below) to a high degree of accuracy.

The unbarred $\P_r$, $r\in\{A_1,A_1^{44},T_1,T_2,E\}$, are
more singular than the barred $\bP_r$.   Using Eqs.~(\ref{vacpolproj})
and~(\ref{irreps}), 
we find that 
\begin{eqnarray}
\label{A1extra}
\P_{A_1}&=&\bP_{A_1}+\frac{1}{q^2}\left(\P_s(0)+\frac{q_4^2{\vec q}\,^2}{q^2(3q^2-{\vec q}\,^2)}\,(\P_4(0)-\P_s(0))\right)\ ,\quad q^2\ne 0\ ,\\
\P_{A_1^{44}}&=&\bP_{A_1^{44}}+
\frac{1}{q^2}\left(\P_s(0)
+\frac{{\vec q}\,^2}{q^2}\,(\P_4(0)-\P_s(0))\right)\ ,\quad {\vec q}\,^2\ne 0\ ,\nonumber\\
\P_{T_1}&=&\bP_{T_1}+\frac{1}{q^2}\left(\P_s(0)
+\frac{{\vec q}\,^2}{q^2}\,(\P_4(0)-\P_s(0))\right)\ ,\quad q_4q_i\ne 0\ \mbox{for\ some}\ i\ ,\nonumber\\
\P_{T_2}&=&\bP_{T_2}+\frac{1}{q^2}\left(\P_s(0)
-\frac{q_4^2}{q^2}\,(\P_4(0)-\P_s(0))\right)\ ,\quad q_iq_j\ne 0\ \mbox{for\ some}\ i,j\ ,\nonumber\\
\P_{E}&=&\bP_{E}+\frac{1}{q^2}\,\left(\P_s(0)
-\frac{q_4^2}{q^2}(\P_4(0)-\P_s(0))
\right)\ ,\quad {\vec q}\,^2\ne 0\ \mbox{and}\ {\vec q}\,^2\ne 3q_i^2\ .\nonumber
\end{eqnarray}
The conditions listed for each case follow from the definitions in Eq.~(\ref{irreps}).
Since both $\bP_r$ and $\P_r$ satisfy the WTIs~(\ref{WTI1}) and~(\ref{WTI2}),
the terms in parentheses in Eq.~(\ref{A1extra}) should satisfy these equations, 
and indeed they do.

One may also define scalar functions $\hat\Pi_r$ as in Eq.~(\ref{irreps}) from the subtracted
vacuum polarization but without the projectors present in Eq.~(\ref{vacpolproj}).
In that case, the subtracted vacuum polarization does not satisfy the WTI,
but this is purely a finite-volume effect.   We have that $\hat\P_r=\P_r$
for $r\in\{T_1,T_2,E\}$, and also that $\hat\P_{A_1^{44}}=\P_{A_1^{44}}$
if $\P_4(0)=0$.   The latter condition is approximately satisfied for the
lattice ensembles we will consider in the sense that $\P_4(0)\ll \P_s(0)$,
as a consequence of the fact that $T\gg L$.   In ChPT we find that
$\bP_{A_1}$ is a smoother function than $\hat\P_{A_1}$, while for the
other representations we find that $\hat\P_r=\P_r$ is smoother than
$\bP_r$.   In Sec.~\ref{lattice}, we will thus choose to consider the lattice data
for $\bP_{A_1}$ and $\P_r$ for $r\in\{A_1^{44},T_1,T_2,E\}$, even though
the difference between $\hat\P_r$ and $\bP_r$ is not visible in the lattice
data because of the statistical errors.

Of course, the data we will consider live not only in a finite volume, but also 
on a lattice.   However, since we are interested in the low-$q$ behavior of 
the vacuum polarization, we will assume that higher-order terms in the
lattice spacing (\seef\ the second line of Eq.~(\ref{vacpolanonzero})) can be
neglected, as mentioned before.   The only scaling violations we will take
into account are those represented by replacing $q_\m$ by $\hq_\m$,
defined in Eq.~(\ref{qhat}), and the taste splitting of the Nambu--Goldstone
boson (NGB) masses present in the spectrum of lattice QCD with staggered fermions
at non-zero lattice spacing.   We note that the numerical difference between
$q_\m$ and $\hq_\m$ is tiny, for momenta up to 1~GeV, for the lattice
ensemble we will consider in this article.

%%####%%
%\newpage
\subsection{\label{ChPT} Chiral perturbation theory}
%%####%%
The heuristic picture is that finite volume
effects are caused by hadrons traveling ``around the world'' (\ie, seeing the
periodic boundary conditions).   The euclidean
propagator for a particle with mass $m$ traveling a distance $L$ falls like $e^{-mL}$.   Therefore, 
finite-volume effects are predominantly felt by the pions, because they are the
lightest hadrons present in the theory, and it is thus useful to
consider finite-volume effects in ChPT, the 
effective field theory for pions.\footnote{See Ref.~\cite{MGLH} for an introduction to applications of ChPT to lattice QCD, including ChPT in a finite volume,
partial quenching, and staggered ChPT, as well as references.} 

It is well-known that leading-order ChPT does not describe the hadronic vacuum
polarization very well already at very low $q^2$ and pion masses \cite{AB2007}.\footnote{For a discussion of $\P_{V-A}$, see Ref.~\cite{L10}.}   The intuitive reason
is that resonance contributions, like that from the $\r$, are important, but 
only higher orders in ChPT ``know'' about such resonances (through low-energy
constants at higher order).   However, by the same argument,
ChPT should do much better describing {\it differences} caused only by 
finite-volume effects, because those should be dominated by pions, and quite
suppressed for all other hadrons.   Here we will assume that it is reasonable
to study finite-volume effects using leading-order ChPT for pions only.   We 
will then compare the predictions from ChPT with lattice data, to see how this
assumption fares, in Sec.~\ref{lattice}.

The lattice data we will consider have been generated on ensembles with three
flavors of sea quarks, up, down and strange.    Therefore, at lowest order
in ChPT, $\P_{\m\n}(q)$ receives loop contributions from all NGBs for a 
three-flavor theory.   However, since the kaon mass is always much larger than the
pion mass, we will calculate only the pion loops in ChPT, and compare the
result with the lattice data.   Any discrepancies may be due to kaon loops, 
higher orders, \etc\ 

The appendix provides some details about the ChPT calculation, and generalizes it
to the case of twisted boundary conditions.   For periodic boundary conditions
the leading-order contribution from pions to the connected part of the vacuum polarization is
\begin{eqnarray}
\label{chptresult}
\P_{\m\n}(q)&=&
\frac{10}{9}\,e^2\,\frac{1}{L^3T}\sum_p\Biggl[
\frac{4\sin{\left(p+q/2\right)_\mu}\sin{\left(p+q/2\right)_\nu}}
{\left(2\sum_\kappa(1-\cos{p_\kappa})+m_\pi^2\right)
\left(2\sum_\kappa(1-\cos{(p+q)_\kappa})+m_\pi^2\right)}\nonumber\\
&&\phantom{\frac{5}{9}\,e^2\,\frac{1}{L^3T}\sum_p\Biggl[}
-\delta_{\mu\nu}\,
\left(\frac{2\cos{p_\mu}}{\left(2\sum_\kappa(1-\cos{p_\kappa)}+m_\pi^2\right)}
\right)\Biggr]\ ,
\end{eqnarray}
where $e$ is the electron charge and $m_\p$ is the pion mass.   We have used
a lattice regulator in order to define this expression, and all dimensionful
physical quantities in Eq.~(\ref{chptresult}) are expressed in units of the lattice
spacing.   It is straightforward to verify that $\P_{\m\n}(0)=0$ 
when the momentum sum in Eq.~(\ref{chptresult}) is replaced by a momentum integral,
by partial integration on the first term in the integral.   In the appendix we show
that in finite volume $\P_{\m\n}(0)\ne 0$, as a simple application of the Poisson
resummation formula, \seef\ Eq.~(\ref{Pi0finvol}). 

Since in the next section we will be comparing ChPT with lattice data obtained
with ``rooted'' staggered fermions, we should amend Eq.~(\ref{chptresult}) to what we would
have obtained using rooted staggered ChPT \cite{LS,AB}.    Staggered fermions lead
to ``taste symmetry breaking,'' splitting the degenerate pion spectrum due to
lattice artifacts, and this turns out to be a signicant effect even at low energy,
and therefore we will take this effect into account when comparing with lattice
data.\footnote{For an introduction to rooted staggered fermions, and further 
references, see Ref.~\cite{MG2008}.}   We will also use the momenta $\hq$ introduced in Eq.~(\ref{qhat}) instead
of $q$, but this amounts to a difference of less that $0.1$\% even at $q^2=1$~GeV$^2$ for the data we consider in Sec.~\ref{lattice}.
 
It is very simple to adapt our result~(\ref{chptresult})
(or Eq.~(\ref{Fresult}) in the appendix) to the staggered case.  To lowest order in rooted staggered ChPT,
all we need to do is to replace the summand in Eq.~(\ref{chptresult}) (or Eq.~(\ref{Fresult}))
by a weighted average over the taste-split pion spectrum with masses
$m_\p=m_P$, $m_A$, $m_T$, $m_V$ and $m_I$, with, respectively,
weights $1/16$, $1/4$, $3/8$, $1/4$, and $1/16$.   We will refer to this
version of our result as (lowest-order) SChPT.

%%####%%
%\newpage
\section{\label{lattice} Lattice data}
%%####%%
In this section, we will consider lattice data for the connected part of the
light-quark hadronic 
vacuum polarization for  the 
asqtad ensemble generated by the MILC collaboration \cite{MILC}
with $1/a=3.35$~GeV, $m_\p=220$~MeV, $L/a=64$ and $T/a=144$.   

For illustration, we show the vacuum polarization for the
asqtad ensemble in Fig.~\ref{asqtadA1A144}, with errors obtained using 
all-mode averaging (AMA) \cite{AMA}.   The red squares show $\bP_{A_1}(\hq^2)$,
obtained from $\bP_{\m\n}(\hq)$ using Eq.~(\ref{irreps}), the blue circles show
$\P_{A_1}(\hq^2)$, and the magenta stars show $\P_{A_1^{44}}(\hq^2)$, obtained similarly from $\P_{\m\n}(\hq)$ (they are slightly offset horizontally to make them
visible).\footnote{The magenta points start at a higher value of $\hq^2$, because
$\P_{A_1^{44}}(\hq^2)$ vanishes when all spatial components of the momentum
vanish, \seef\ Eq.~(\ref{irreps}).  In contrast, $\bP_{A_1}(\hq^2)$ and $\P_{A_1}(\hq^2)$ do not vanish for any non-zero value of the momentum.}   The difference 
between these three cases is a finite-volume effect, and clearly visible, thanks
to the very small statistical errors obtained with the AMA method.   The lattice
data for $\bP_{A_1^{44}}(\hq^2)$ agree, within errors, with those for 
$\P_{A_1^{44}}(\hq^2)$.  

%%%%%%%%%%%%%%%%%%%
\begin{figure}[t!]
%\vspace*{4ex}
\begin{center}
\includegraphics*[width=12cm]{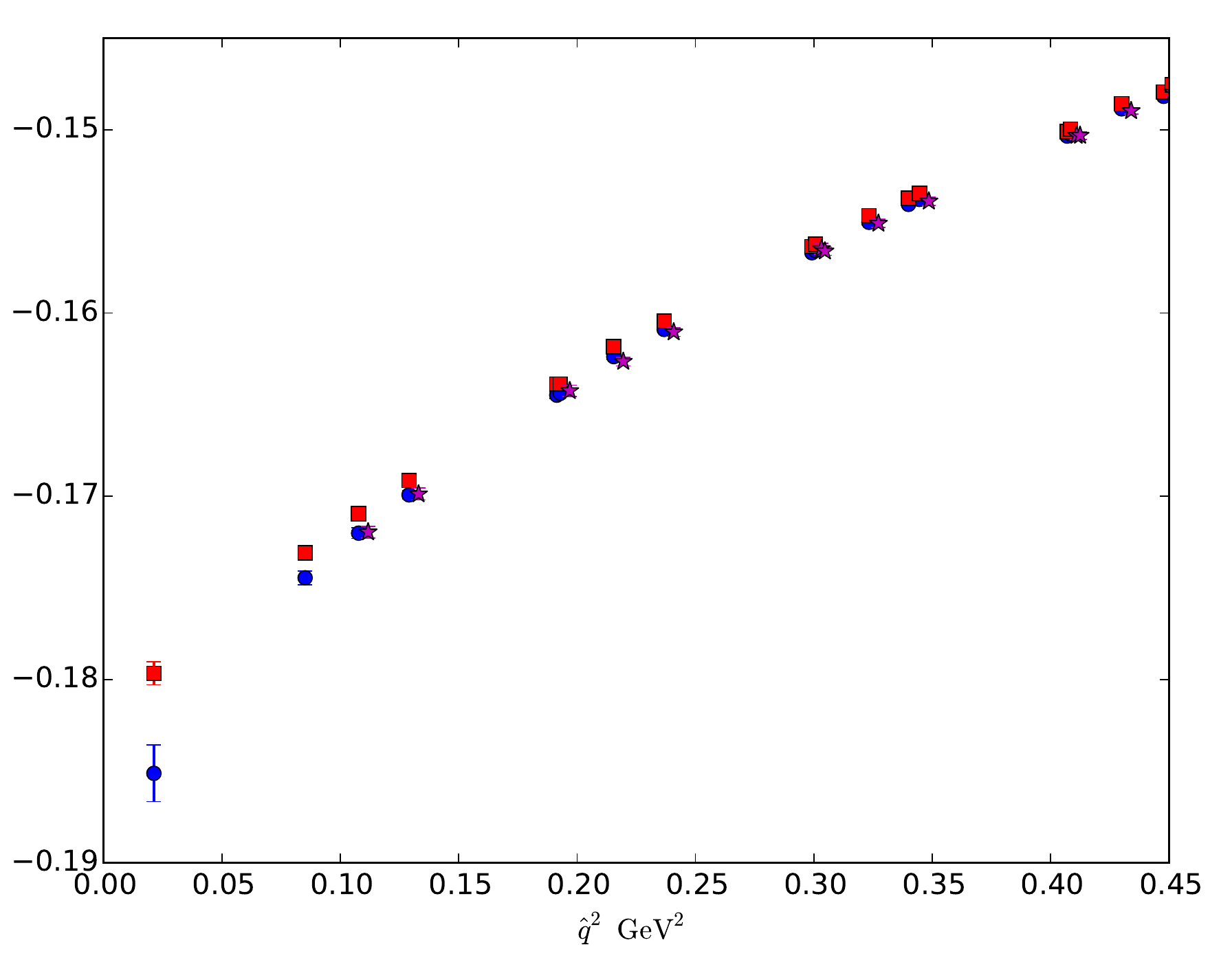}
\end{center}
\vspace*{-6ex}
\begin{quotation}
\floatcaption{asqtadA1A144}%
{{\it Low-$\hq^2$ lattice data for the connected part of $\bP_{A_1}(\hq^2)$ (red squares), $\P_{A_1}(\hq^2)$ (blue circles), and $\P_{A_1^{44}}(\hq^2)$ (purple stars).  
MILC asqtad ensemble; the purple stars have been
horizontally offset by $+0.004$~{\rm GeV}$^2$ for visibility.  }}
\end{quotation}
\vspace*{-4ex}
\end{figure}
%%%%%%%%%%%%%%%%%%%

We will first compare the lattice data to ChPT, in Sec.~\ref{comparison}.  Since, as explained in Sec.~\ref{ChPT}, 
we expect that only finite-volume differences can be sensibly compared, we will
limit ourselves to such differences.   Then, in Sec.~\ref{amuvalues}, we will use the lattice data to 
determine $a_\m[\hq^2_{max}=0.1\ \mbox{GeV}^2]$ from different irreps, in order to see how
finite-volume effects in the data propagate to the anomalous magnetic moment.

%%####%%
%\newpage
\subsection{\label{comparison} Comparison with ChPT}
%%####%%
%%%%%%%%%%%%%%%%%%%
\begin{figure}[t]
%\vspace*{4ex}
\begin{center}
\includegraphics*[width=12cm]{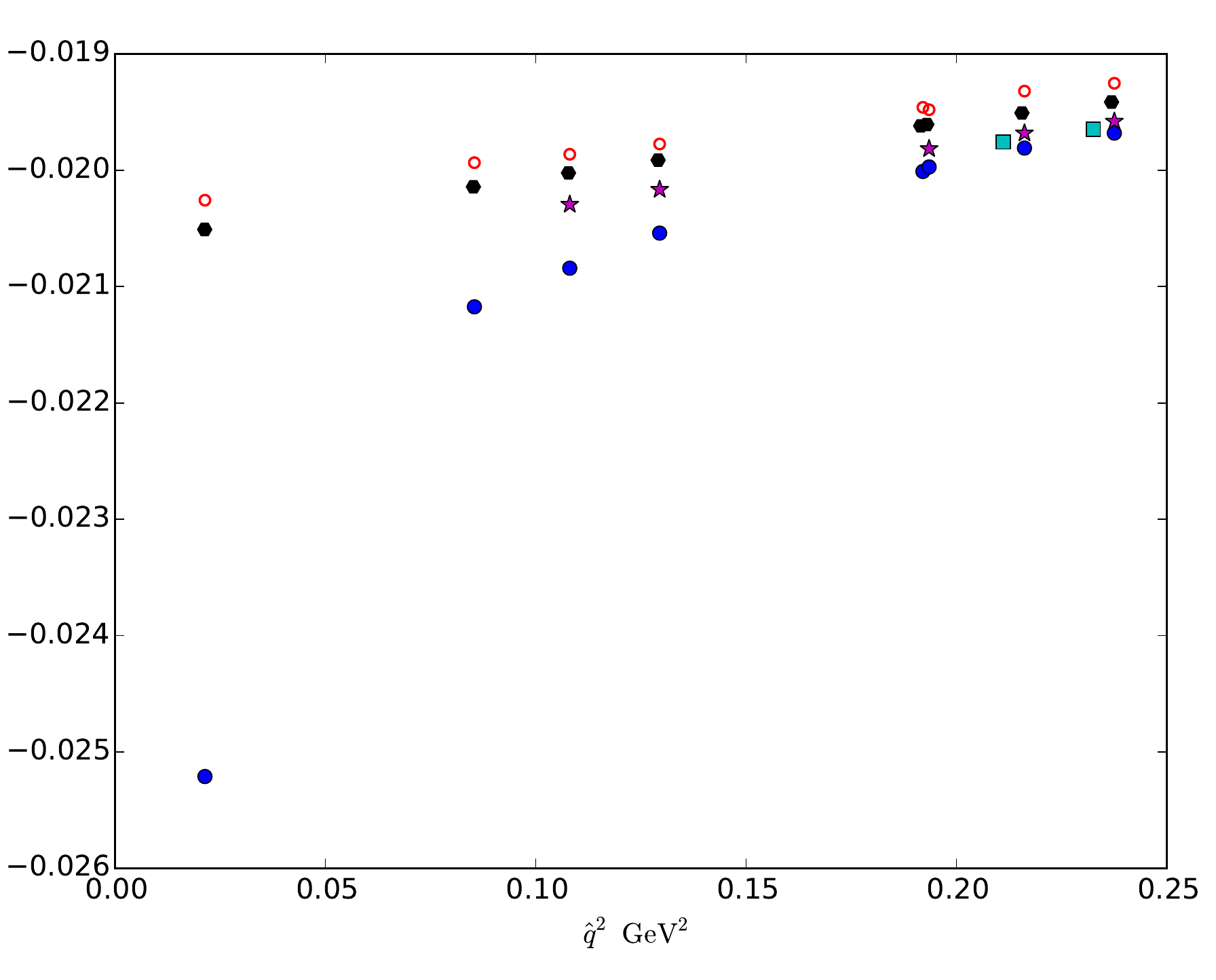}
\end{center}
\vspace*{-6ex}
\begin{quotation}
\floatcaption{CHPTA1A144}%
{{\it Low-$\hq^2$ SChPT points for the connected part of
$\bP_{A_1}(\hq^2)$ (red open circles), $\P_{A_1}(\hq^2)$ (blue filled circles), $\P_{A_1^{44}}(\hq^2)$ (purple stars), $\P_{T_2}(\hq^2)$
(cyan squares), and $\P_{A_1}(\hq^2)$ in
infinite volume (black hexagons).  }}
\end{quotation}
\vspace*{-4ex}
\end{figure}
%%%%%%%%%%%%%%%%%%%
Figure~\ref{CHPTA1A144} shows a plot similar to Fig.~\ref{asqtadA1A144}, but
with the data points computed in lowest-order SChPT.
In addition, in ChPT we have access to the values of the vacuum polarization
in infinite volume, and infinite-volume points for the same $\hq^2$ values are shown in black
in Fig.~\ref{CHPTA1A144}.\footnote{In reality, the black points are $\P_{A_1}(\hq^2)$
for $L/a=128$, $T/a=288$.   On the scale of the plot, the differences between
infinite-volume points and the black points, or the differences between 
$\bP_r(\hq^2)$ and $\P_r(\hq^2)$, $r\in\{A_1,A_1^{44},T_1,T_2,E\}$, are not 
visible.   We always omit a factor $5e^2/9$, equal to the sum of the 
squares of the charges
of the up and down quarks,
from Eq.~(\ref{chptresult}), and we do the same for the
lattice data.} 

 Consistent with what we observe in the lattice
data, there is no significant difference between $\bP_r(\hq^2)$ and $\P_r(\hq^2)$
for all representations except $r=A_1$.   This is why we omitted $\bP_{A_1^{44}}(\hq^2)$ and the representations $T_1$ and $E$ in Fig.~\ref{CHPTA1A144}.  We
do show $\P_{T_2}(\hq^2)$ as the two cyan squares all the way to the right.\footnote{
We recall also that only three out of the five $\P_r$ and $\bP_r$ are independent, because of the
relations~(\ref{WTI1}) and~(\ref{WTI2}).}
To extract $\P_{T_2}(\hq^2)$ from $\P_{ij}(\hq)$ we need two different spatial
components of the momentum to not vanish, implying that $\hq^2\ge 8\p^2/L^2
=0.216$~GeV$^2$ for these points.   
The $A_1$ representation is the most interesting case, because it reaches lower
momenta than any of the others. It is the only one that does not vanish when
only $\hq_4\ne 0$, and the two smallest values of $\hq^2$ have only $\hq_4\ne 0$;
it is thus the most useful representation to explore the low-momentum behavior of
$\P(q^2)$.

There are clear differences between Figs.~\ref{asqtadA1A144} and\ \ref{CHPTA1A144}.   First, the vertical offset is very different.   This is not a
physical effect, because the quantities plotted are logarithmically divergent in
the continuum limit.   However, the slopes are also vastly different, and this is
a physical effect, already observed in Ref.~\cite{AB2007}.   The slope of the
vacuum polarization at low $q^2$ is dominated by the $\r$ resonance, but this resonance (and
others) are absent in Eq.~(\ref{chptresult}).\footnote{This observation of Ref.~\cite{AB2007}
has led to the ubiquitous use of vector-meson dominance to parametrize the
vacuum polarization, before model-independent methods started to be explored
\cite{taumodel,Pade,strategy,HPQCD}.}

Despite these differences, there are useful lessons to be learned from Fig.~\ref{CHPTA1A144}.   The subtracted value $\bP_{A_1}(\hq^2)$ is an order
of magnitude closer to the infinite-volume points than the unsubtracted value,
$\P_{A_1}(\hq^2)$.   Clearly, the lesson is that one should carry out the
subtraction~(\ref{vacpolproj}) (at least for the $A_1$ representation).   This was already observed empirically in
Ref.~\cite{BMW}, and we see here that this observation is theoretically supported by
 ChPT.
Furthermore, we see that $\bP_{A_1}(\hq^2)$ and $\P_{A_1^{44}}(\hq^2)$
straddle the infinite-volume result, suggesting that also in lattice QCD
the true value of $\P(q^2)$ lies in between these two.\footnote{$\P_r(\hq^2)$
for $r\in\{T_1,T_2,E\}$ also lies below the infinite-volume result close to
$\P_{A_1^{44}}(\hq^2)$, according to ChPT.}

Of course, one would like to test whether these lessons from lowest-order
SChPT also apply to the actual lattice data.   While no lattice data are 
available in infinite volume, it is possible to compare finite-volume differences
predicted by SChPT to such differences computed from the lattice data.
%%%%%%%%%%%%%%%%%%%
\begin{figure}[t]
%\vspace*{4ex}
\begin{center}
\includegraphics*[width=12cm]{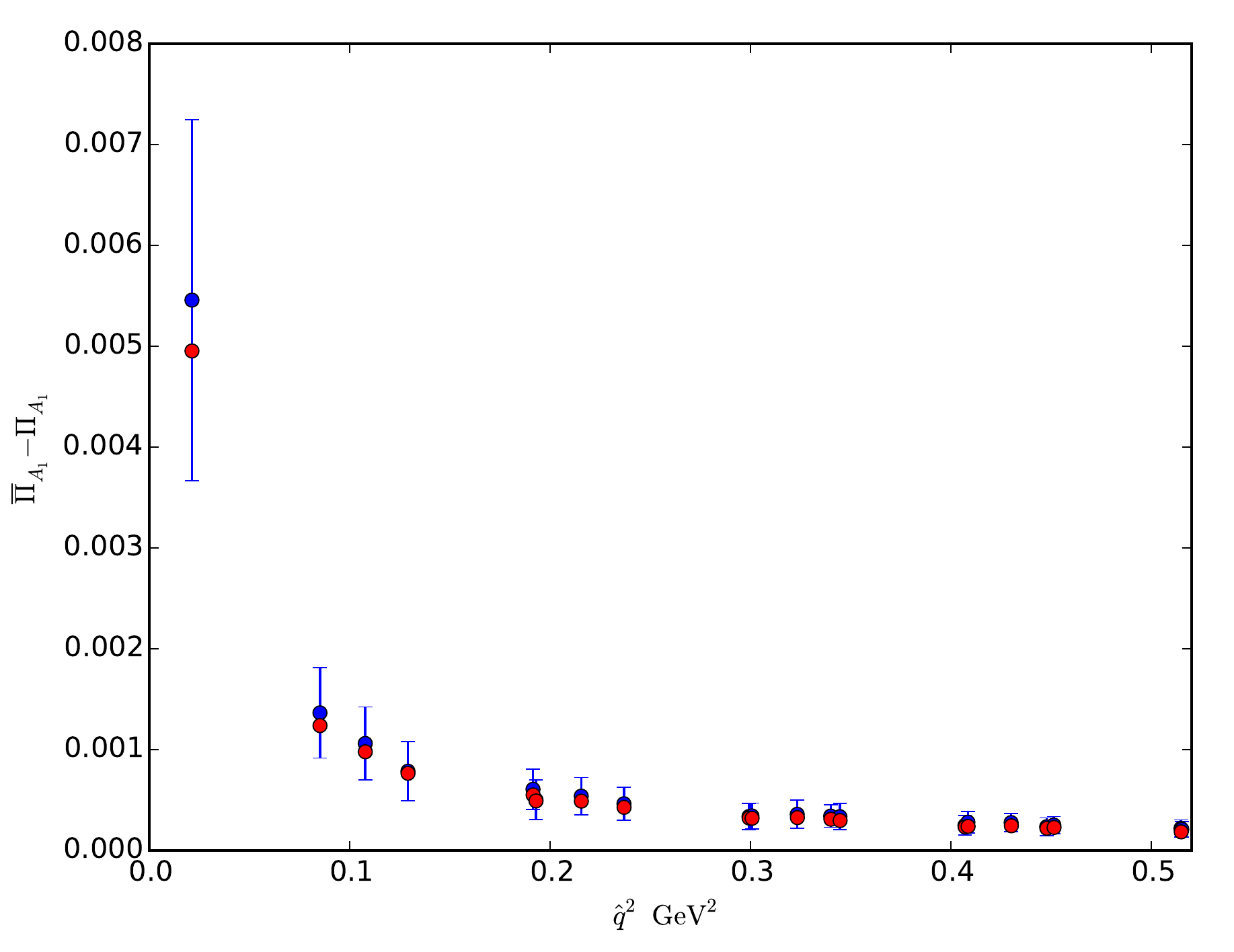}
\end{center}
\vspace*{-6ex}
\begin{quotation}
\floatcaption{A1diffchptvsasqtad}%
{{\it Comparison of $\bP_{A_1}(\hq^2)-\P_{A_1}(\hq^2)$ between MILC asqtad
lattice data
(blue points) and lowest-order SChPT (red points).}}
\end{quotation}
\vspace*{-4ex}
\end{figure}
%%%%%%%%%%%%%%%%%%%
In Fig.~\ref{A1diffchptvsasqtad} we show the difference  $\bP_{A_1}(\hq^2)-\P_{A_1}(\hq^2)$ in the low-$\hq^2$ region, both on the lattice and computed in lowest-order
SChPT.   This difference is a pure finite-volume effect.   Clearly, SChPT does a
very good job of describing the lattice data, with all red points 
within less than $1\s$ of the blue points. 
This is remarkable, especially in view of the fact that
lowest-order SChPT does such a poor job of describing the full lattice data for
$\P_{A_1}(\hq^2)$, as we noted above.  

%%%%%%%%%%%%%%%%%%%
\begin{figure}[t]
%\vspace*{4ex}
\begin{center}
\includegraphics*[width=12cm]{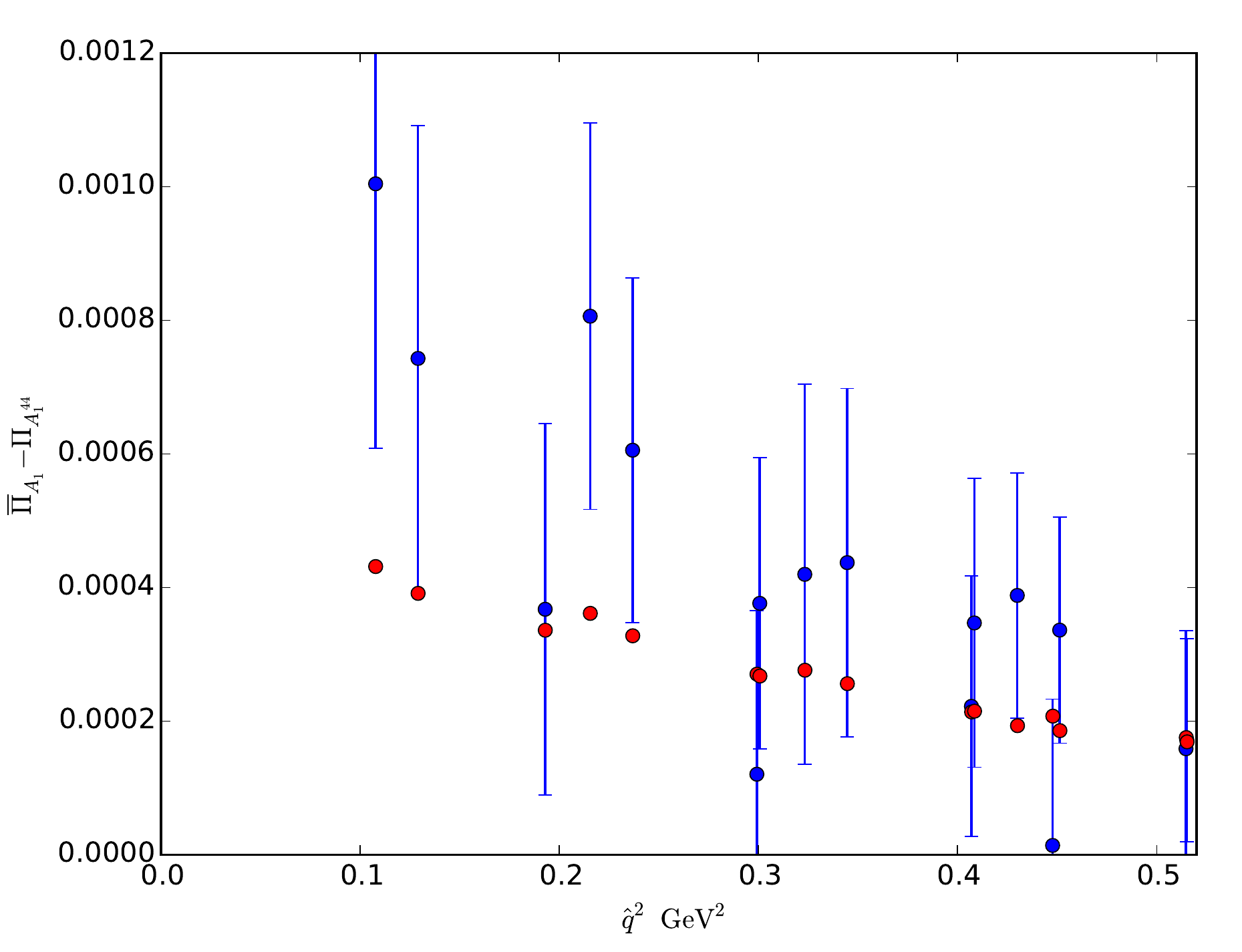}
\end{center}
\vspace*{-6ex}
\begin{quotation}
\floatcaption{A1subA144unsubdiffchptvsasqtad}%
{{\it Comparison of $\bP_{A_1}(\hq^2)-\P_{A_1^{44}}(\hq^2)$ between MILC asqtad lattice data
(blue points) and lowest-order SChPT (red points).}}
\end{quotation}
\vspace*{-4ex}
\end{figure}
%%%%%%%%%%%%%%%%%%%
We may also consider differences between different representations, which 
also probes the size of finite-volume effects.   In Fig.~\ref{A1subA144unsubdiffchptvsasqtad} we show the difference $\bP_{A_1}(\hq^2)-
\P_{A_1^{44}}(\hq^2)$, for the lattice data, and computed in SChPT.   
To extract $\P_{A_1^{44}}(\hq^2)$ from $\P_{\m\n}(\hq)$ we need at least one spatial
component of the momentum to not vanish, implying that $\hq^2\ge 4\p^2/L^2
=0.108$~GeV$^2$ for these points. 
All
observations made above about the difference $\bP_{A_1}(\hq^2)-
\P_{A_1}(\hq^2)$ apply here as well, with the difference between lattice data and ChPT now averaging about $1\s$.  We note the difference of scale on the
vertical axis between Figs.~\ref{A1diffchptvsasqtad} and \ref{A1subA144unsubdiffchptvsasqtad}, consistent with the fact that both $\bP_{A_1}(\hq^2)$ and
$\P_{A_1^{44}}(\hq^2)$ are much closer to the infinite-volume limit than
$\P_{A_1}(\hq^2)$.   We find that the pattern is very similar for other
representations.    

%%####%%
%\newpage
\begin{boldmath}
\subsection{\label{amuvalues} Effects on $a_\m^{\rm HVP}$}
\end{boldmath}
%%####%%
Finally, while it is already clear that there are significant finite-volume effects
in the vacuum polarization, we consider the question of how they propagate
to the anomalous magnetic moment itself.    We will, in fact, compare the
quantity $a_\m^{\rm LO,HVP}[\hq_{max}^2]$ with the choice $\hq_{max}^2=0.1$~GeV$^2$, in order to be certain that differences are due to finite volume,
and not to lattice spacing effects.\footnote{More than 80\% of $a_\m^{\rm LO,HVP}$
comes from the momentum region below $0.1$~GeV$^2$ \cite{strategy}.}

We fit the data for $\bP_{A_1}$ and $\P_{A_1^{44}}$ with a $[0,1]$ Pad\'e \cite{Pade}, or a quadratic conformally mapped polynomial \cite{strategy} (both are three-parameter
fits), on a low-$q^2$ interval, looking for the number of data points in the fit
that gives the highest $p$-value.
We then compare the results.   In all the fits presented below, the number of data
points in the fit turns out to be six, so all fits have three degrees of freedom,
and they never explore data beyond $\hq^2=0.3$~GeV$^2$.
As we have shown before \cite{taumodel,strategy},
neither of these two fits can be trusted to give results with a better accuracy than
a few percent even in infinite volume for $a_\m^{\rm LO,HVP}$, but we will assume that other systematics are the same for both 
representations, so that the differences considered here measure primarily finite-volume effects.
From the $[0,1]$ Pad\'e fits, we find
\begin{eqnarray}
\label{Pade}
a_{\m,A_1}^{\rm LO,HVP}[0.1\ \mbox{GeV}^2]&=&6.8(4)\times 10^{-8}\ ,\\
a_{\m,A_1^{44}}^{\rm LO,HVP}[0.1\ \mbox{GeV}^2]&=&7.5(3)\times 10^{-8}\ .\nonumber
\end{eqnarray}
From the quadratic conformally mapped polynomial fits, we find
\begin{eqnarray}
\label{conformal}
a_{\m,A_1}^{\rm LO,HVP}[0.1\ \mbox{GeV}^2]&=&6.8(4)\times 10^{-8}\ ,\\
a_{\m,A_1^{44}}^{\rm LO,HVP}[0.1\ \mbox{GeV}^2]&=&7.9(4)\times 10^{-8}\ .\nonumber
\end{eqnarray}
Both types of fit give consistent results for each representation, but the two
different representations differ from each other by about 10-15\%.   
This strongly suggests that with a pion mass of $220$~MeV a spatial
volume with $L=64a=3.8$~fm, or $m_\p L=4.2$, is not large enough if the aim
is to compute $a_\m^{\rm LO,HVP}$ with sub-percent accuracy.

\vskip0.8cm
%%####%%
%\newpage
\section{\label{conclusion} Conclusion}
%%####%%
In this article, we explored finite-volume effects in the connected part of the hadronic vacuum 
polarization, and gave some examples of how these effects propagate to
the corresponding contribution to the muon anomalous
magnetic moment $a_\m^{\rm LO,HVP}$.   We found that even in computations 
with small pion masses and  $m_\p L>4$, the systematic 
effects due to finite volume can be of order 10\%.   This is consistent with the
phenomenological estimate of Ref.~\cite{Mainz}.

We also found that ChPT does a good job of describing finite-volume
effects already at lowest order, even though it is well known that lowest-order
does not provide a good description of the vacuum polarization itself already
at the low values of $q^2$ relevant for $a_\m^{\rm LO,HVP}$.   ChPT also
shows that the subtracted vacuum polarization $\bP_{\m\n}(q)=
\P_{\m\n}(q)-\P_{\m\n}(0)$ is significantly closer to the infinite-volume result
than $\P_{\m\n}(q)$ itself.   
Projecting 
on irreducible representations of the cubic group, we found that in ChPT
the $A_1$
projection (after subtraction of $\P_{\m\n}(0)$) and other representations
(for which the subtraction makes very little difference, and is not visible in
the lattice data)
straddle the infinite-volume result.  This leads to the question of how to
quantify the systematic error due to finite volume in practice.   A conservative
error estimate would take half the difference between the value of
$a_\m^{\rm LO,HVP}$ computed from $\bP_{A_1}$ and the values 
computed from other representations, \eg\ $\P_{A_1^{44}}$.   Because
the infinite-volume result for $\P$, according to ChPT, lies between 
$\bP_{A_1}$ and $\P_{A_1^{44}}$, a more aggressive error estimate
would be obtained by taking the difference between the average of the
ChPT results for $\bP_{A_1}$ and $\P_{A_1^{44}}$ and the infinite-volume
ChPT result.    However, this might be too aggressive, because the 
comparison of finite-volume differences between the lattice and ChPT
shows that the (lowest-order) ChPT estimates the finite-volume effects 
we see on the lattice to about $1\s$ (\seef\ Figs.~\ref{A1diffchptvsasqtad} and \ref{A1subA144unsubdiffchptvsasqtad}). For lattice data with increased statistical precision, it is
not clear whether lowest-order ChPT would be precise enough, in particular for
the representation $A_1^{44}$.
It will be interesting to compare the finite-volume effects for physical pion mass and larger volume ($L>5$ fm). Until then it is not advisable to use ChPT to correct the lattice results.

Finally, we wish to make a comment on the ``moment method,'' proposed
in Ref.~\cite{HPQCD}.   In infinite volume, $\P(0)$ can be obtained from the
second moment of the current two-point function (no sum over $i$),
\begin{equation}
\label{moment}
\P(0)=-\half\,\int dt\int d^3{\vec x}\,t^2\,\langle J_i({\vec x},t)J_i({\vec 0},0)\rangle\ .
\end{equation}
In a finite volume $L^3\times T$, using 
\begin{eqnarray}
\label{tsqFT}
t^2&=&\sum_n a_n\,\cos{(2\p nt/T)}\ ,\\
a_0&=&\frac{T^2}{12}\ ,\quad 
a_n=\frac{(-1)^n}{\sin^2{(\p n/T)}}\ ,\quad n\ne 0\ ,\nonumber
\end{eqnarray}
the right-hand side of Eq.~(\ref{moment}) gets replaced by the expression \cite{BI}
\begin{equation}
\label{momfinvol}
\P(0)\to 4\sum_{n\ne 0}(-1)^n \P\left((2\p n/T)^2\right)\ .
\end{equation}
We see that in finite volume, $\P(0)$ gets replaced by a linear combination
of values of $\P(q^2)$ at non-zero values of $q^2$, not including $q^2=0$.
This appears to imply that the moment method is equally susceptible to
finite-volume effects.

\vspace{3ex}
%\newpage
\noindent {\bf Acknowledgments}
\vspace{3ex}

We thank Taku Izubuchi and Kim Maltman for useful discussions, and Luchang Jin for pointing out an error in the first version of this article.
We also thank USQCD for the computing resources used to generate the vacuum polarization as well as the MILC collaboration for providing the configurations used.
TB, PC and MG were supported in part by the US Department of Energy under
Grant No. DE-FG02-92ER40716 and Grant No. DE-FG03-92ER40711.
SP is supported by CICYTFEDER-FPA2014-55613-P, 2014-SGR-1450,
the Spanish Consolider-Ingenio 2010 Program
CPAN (CSD2007-00042).

%%####%%
%\newpage
\appendix
\section{\label{one-loop} Vacuum polarization at one loop in ChPT}
%%####%%
In this appendix, we derive a generalization of Eq.~(\ref{chptresult}) for the case of
twisted boundary conditions.   This is a partially quenched calculation, because we
will only give the valence quarks, \ie, the quarks to which the external photons
couple, a twist, while all sea quarks obey periodic boundary conditions.  We follow
the definitions and conventions of Ref.~\cite{ABGP13}.\footnote{See Ref.~\cite{DJJW2012,PB,GDRPNT,CSGV} for earlier work on twisted boundary conditions for valence quarks.}  Throughout this appendix, we will use the lattice as a UV
regulator, and we will express all quantities in terms of lattice units.

We introduce six quarks,
\begin{equation}
\label{quarkbasis}
q=\left(\begin{array}{c}
u_v \\ u_t \\ d_v \\ d_t \\ u_s \\ d_s \end{array}\right)\ ,
\end{equation}
where the index $v$ labels the untwisted valence quarks, the index $t$ labels the twisted valence quarks and the index $s$ labels the (untwisted) sea quarks.  
The twisted quarks have boundary conditions
\begin{eqnarray}
\label{tbc}
q_t(x)&=&e^{-i\theta_\mu}q_t(x+L_\mu)\ ,\\
\qb_t(x)&=&\qb_t(x+L_\mu)e^{i\theta_\mu}\ ,\nonumber
\end{eqnarray}
with $L_i=L,\ i=1,2,3$ and $L_4=T$.
Strictly speaking, one should also introduce four ghost quarks to cancel loops
of the four valence quarks, but we will leave this implicit in the rest of our
calculation.\footnote{In this calculation it is not difficult to match pion loops with quark loops, so it is easy to identify contributions that should be omitted so as to suppress
valence quark loops.}
Only valence quarks couple to photons, and this coupling takes the form
\begin{equation}
\label{quarkcoupling}
e\qb\g_\m (A_\m^+ Q^++A_\m^- Q^-)q\ ,
\end{equation}
with
\begin{equation}
\label{Qs}
Q^+=\left(\begin{array}{cccccc}
0 & 2/3 & 0 & 0 & 0 & 0\\ 0 & 0 & 0 & 0 & 0 & 0\\ 0 & 0 & 0 & -1/3 & 0 & 0\\ 0 & 0 & 0 & 0 & 0 & 0 \\  0 & 0 & 0 & 0 & 0 & 0 \\  0 & 0 & 0 & 0 & 0 & 0\end{array}\right)\ ,\quad
Q^-=\left(\begin{array}{cccccc}
0 & 0 & 0 & 0 & 0 & 0 \\ 2/3 & 0 & 0 & 0 & 0 & 0\\ 0 & 0 & 0 & 0 & 0 & 0\\ 0 & 0 & -1/3 & 0 & 0 & 0 \\ 0 & 0 & 0 & 0 & 0 & 0 \\  0 & 0 & 0 & 0 & 0 & 0\end{array}\right)\ .
\end{equation}
In order to accommodate the twist, there are {\it two} photons, one coupling to
the current $\frac{2}{3}\ubar_t\g_\m u_v-\frac{1}{3}\dbar_t\g_\m d_v$, and 
one coupling to the current $\frac{2}{3}\ubar_v\g_\m u_t-\frac{1}{3}\dbar_v\g_\m d_t$,
corresponding to the photons going into, and out of, the vacuum polarization,
which, at the valence-quark level, consists of a loop made out of a twisted
up or down quark and an untwisted up or down anti-quark, thus inserting a
momentum $q_\m+\theta_\m/L_\m$
with $q$ a periodic momentum as in Eq.~(\ref{momquant}),
and $\theta_\m\in[0,2\p)$.

In this theory with six quarks,\footnote{And four ghost quarks.} the pions form a $6\times 6$ matrix, with flavor structure\footnote{This equation was incorrect in the first version
of this paper.  This led to Eq.~(\ref{Fresult}) to be a factor 2 too small in the first
version.   We thank
Luchang Jin for pointing this out to us.}
\begin{equation}
\label{pionfield}
\phi\sim\left(\begin{array}{cccccc}
u_v\ubar_v & u_v\ubar_t & u_v\dbar_v & u_v\dbar_t & u_v\ubar_s & u_v\dbar_s\\ u_t\ubar_v & u_t\ubar_t & u_t\dbar_v & u_t\dbar_t & u_t\ubar_s & u_t\dbar_s\\ d_v\ubar_v & d_v\ubar_t & d_v\dbar_v & d_v\dbar_t & d_v\ubar_s & d_v\dbar_s\\ d_t\ubar_v & d_t\ubar_t & d_t\dbar_v & d_t\dbar_t
& d_t\ubar_s & d_t\dbar_s\\ u_s\ubar_v & u_s\ubar_t & u_s\dbar_v & u_s\dbar_t & u_s\ubar_s & u_s\dbar_s \\ d_s\ubar_v & d_s\ubar_t & d_s\dbar_v & d_s\dbar_t & d_s\ubar_s & d_s\dbar_s\end{array}\right)
\sim
\left(\begin{array}{cccccc}
 &  &  &   & \pi^{uu}_{vs} &  \pi^{ud}_{vs}\\
   &  &   &   & \pi^{uu}_{ts} &  \pi^{ud}_{ts}\\
   &  &  &  & \pi^{du}_{vs} & \pi^{dd}_{vs}\\
    &  &  & & \pi^{du}_{ts} & \pi^{dd}_{ts}\\
     \pi^{uu}_{sv}    & \pi^{uu}_{st} & \pi^{ud}_{sv} & \pi^{ud}_{st} &  &  \\
         \pi^{du}_{sv} & \pi^{du}_{st} & \pi^{dd}_{sv} & \pi^{dd}_{st} &  & \end{array}\right)\ ,
\end{equation}
where in the second expression we omitted all pions that do not contribute, and we used a superscript to 
indicate the up/down flavor structure.   Note that for instance $\pi^{uu}_{vs}$ is a
charged pion, because it consists of a valence up quark with charge $2/3$, and a
neutral up sea anti-quark.   Pure sea pions do not contribute because they are
neutral.

Pions with no $t$ subscript or with a $tt$ subscript have periodic 
boundary conditions, but pions with one $t$ subscript inherit twisted
boundary conditions from Eq.~(\ref{tbc}), for example ($i,\ j=u,\ d$)
\begin{equation}
\label{twpion}
\pi^{ij}_{ts}(x+L_\m)=e^{i\theta_\m}\pi^{ij}_{ts}(x)\ ,\qquad \pi^{ij}_{st}(x+L_\m)=e^{-i\theta_\m}\pi^{ij}_{st}(x)\ .
\end{equation}
To leading order in ChPT, our calculation is equivalent to a scalar QED calculation,
with lagrangian
\begin{equation}
\label{lagr}
{\cal L}=\frac{1}{2}\,\mbox{tr}\left(D_\m\phi D_\m\phi\right)\ ,
\end{equation}
with a covariant derivative accommodating the gauge invariance implied
by Eq.~(\ref{quarkcoupling}),
\begin{equation}
\label{covder}
D_\m\phi=\partial_\m\phi+ie[A_\m^+ Q^+ + A_\m^- Q^-,\phi]\ .
\end{equation}
It is now straightforward to carry out the desired one-loop calculation, in which
the twisted vacuum polarization is defined as 
\begin{equation}
\label{tPi}
\P^{+-}_{\m\n}(x-y)=\frac{\partial^2}{\partial A_\m^+(x)\partial A_\n^-(y)}\log{Z}
\equiv e^{i\htheta(x+\m/2-y-\n/2)}F_{\mu\nu}^{+-}(x-y)\\ ,
\end{equation}
with $Z$ the path integral with lagrangian~(\ref{lagr}), and
$\htheta_\m=\theta_\m/L_\m$.   Using the lattice as a regulator
by replacing the covariant derivative~(\ref{covder}) by the nearest-neighbor
covariant derivative
\begin{equation}
\label{latcovder}
D_\m\phi(x)=U_\m(x)\phi(x+\m)U_\m^\dagger(x)-\phi(x)\ ,
\end{equation}
with $U_\m(x)=\mbox{exp}[ie(A_\m^+(x)Q^++A_\m^-(x)Q^-)]$,
we find the result generalizing Eq.~(\ref{chptresult}) for non-zero twist:
\begin{eqnarray}
\label{Fresult}
F_{\mu\nu}^{+-}(k)&=&\\
&&\hspace{-2.2cm}\frac{10}{9}\,e^2\,\frac{1}{L^3T}\sum_p\Biggl[
\frac{4\sin{\left(p+(k+\htheta)/2\right)_\mu}\sin{\left(p+(k+\htheta)/2\right)_\nu}}
{\left(2\sum_\kappa(1-\cos{p_\kappa})+m_\pi^2\right)
\left(2\sum_\kappa(1-\cos{(p+k+\htheta)_\kappa})+m_\pi^2\right)}\nonumber\\
&&\hspace{-2.4cm}\phantom{\frac{5}{9}\,e^2\,\frac{1}{L^3T}\sum_p}-\delta_{\mu\nu}\,
\Biggl(\frac{\cos{p_\mu}}{\left(2\sum_\kappa(1-\cos{p_\kappa)}+m_\pi^2\right)}
+\frac{\cos{(p+\htheta)_\mu}}{\left(2\sum_\kappa(1-\cos{(p+\htheta)_\kappa)}+m_\pi^2\right)}\Biggr)\Biggr]\ .
\nonumber
\end{eqnarray}
In the infinite-volume limit, this result agrees with that of Ref.~\cite{DJ}.

It is easy to verify that for $L,T\to\infty$, $F^{+-}_{\m\n}(0)=0$.   The sum in
Eq.~(\ref{Fresult}) becomes an integral, $\htheta\to 0$, and we can partially integrate
the first term to cancel against the second.   But, in a finite volume, these 
simplifications do not apply.   For zero twist, $\theta_\m=0$, it is straightforward
to estimate $F^{+-}_{\m\n}(0)=\P_{\m\n}(0)$ analytically.   Only the diagonal terms
do not vanish, and $\P_{ii}(0)\gg \P_{44}(0)$ if $T\gg L$.   We will therefore
choose $\m=\n=1$.
Using 
\begin{equation}
\label{perdelta}
\sum_{n=-\infty}^\infty\d(x-n)=\sum_{n=-\infty}^\infty e^{2\p i nx}\ ,
\end{equation}
and denoting the expression inside square brackets in Eq.~(\ref{Fresult}) for $\theta_\m=0$ and $q=0$ as $f_{\m\n}(p)$, Eq.~(\ref{Fresult}) can be rewritten as (dropping the factor $10e^2/9$)
\begin{equation}
\label{step1}
\frac{1}{L^3T}\sum_{p}f_{11}(p)=\frac{1}{(2\p)^4}\sum_n \int d^4p f_{11}(p)\,e^{inLp}\ ,
\end{equation}
where $nLp=\sum_\m n_\m L_\m p_\m$.   The term with $n=0$ is the infinite volume
result, and the terms with $n\in\{(\pm 1,0,0,0),(0,\pm 1,0,0),(0,0,\pm 1,0)\}$ 
constitute the dominant finite-volume correction.   Focussing on these terms,
we can take the continuum limit, yielding the intermediate result
\begin{equation}
\label{step2}
\frac{1}{(2\p)^4}\int d^4p\left(\frac{4p_1^2}{(p^2+m_\p^2)^2}-\frac{2}{p^2+m_\p^2}\right)
\left(2e^{iLp_1}+4e^{iLp_2}\right)\ .
\end{equation}
Carrying out the integral over $p_1$ we find that the
integral with the factor $e^{iLp_2}$ vanishes (by partial integration on $p_1$ of the first
term), and
this expression reduces to 
\begin{equation}
\label{step3}
-2L\,\frac{1}{(2\p)^3}\int d^3p\,e^{-L\sqrt{m_\p^2+{\vec p}^2}}=-\frac{m_\p^2}{\p^2}\,
K_2(m_\p L)\ ,
\end{equation}
where $K_2(z)$ is a modified Bessel function of the second kind.   Using its
asymptotic expansion for large argument, we find that (\seef\ Eq.~(\ref{formPi0}))
\begin{equation}
\label{Pi0finvol}
\P_s(0)=\P_{11}(0)\sim -\frac{10e^2}{9}\frac{m_\p^2}{\p^2}\sqrt{\frac{\p}{2m_\p L}}\,e^{-m_\p L}\left(1+
O\left(\frac{1}{m_\p L}\right)\right)\ .
\end{equation}

%%%%%%%%%%%%%%%%%%%%%%%%%%%%%%%%%%%%%%%%%%%%%%%%%%%%%%%%%%%%%%%
%\newpage

\end{document}